\documentclass[10pt,english]{article}
\usepackage{times}
\usepackage[T1]{fontenc}
\usepackage[latin1]{inputenc}
\usepackage{amsmath}
\usepackage{amssymb}

\makeatletter

\providecommand{\tabularnewline}{\\}

\newcommand{\tensor}[1]{\stackrel{\leftrightarrow}{#1}}

\usepackage{slashed}

\usepackage{babel}
\makeatother
\begin{document}

\title{\textbf{Electro-weak $SU(4)_{L}\otimes U(1)_{Y}$ models without
exotic electric charges}}

\author{ADRIAN PALCU}

\date{\emph{Faculty of Exact Sciences - {}``Aurel Vlaicu'' University
Arad, Str. Elena Dr\u{a}goi 2, Arad - 310330, Romania}}

\maketitle
\begin{abstract}
For the particular class of \textbf{$SU(4)_{L}\otimes U(1)_{Y}$}
electro-weak models without exotic electric charges, some plausible
phenomenological predictions - such as the boson mass spectrum and
charges of all the fermions involved therein - are made by using the
algebraical approach of the exactly solving method for gauge models
with high symmetries. Along with the one-parameter resulting mass
scale (to be confirmed at TeV scale in LHC, LEP, CDF and other high
energy experiments) our approach predicts the exact expressions of
the charges (both electric and neutral) in the fermion sector, while
all the Standard Model phenomenology is naturally recovered. 


\end{abstract}

\section{Introduction}

In view of new experimental challenges - such as tiny massive neutrinos
and their oscillations or extra-neutral gauge bosons, to mention but
a few - the Standard Model (SM) \cite{key-1} - based on the gauge
group $SU(3)_{C}\otimes SU(2)_{L}\otimes U(1)_{Y}$ that undergoes
in its electro-weak sector a spontaneous symmetry breakdown (SSB)
up to the electromagnetic universal one $U(1)_{em}$ - has to be properly
extended. One of the most appealing extensions of the SM relies on
the gauge symmetry $SU(3)_{C}\otimes SU(4)_{L}\otimes U(1)_{Y}$ (hereafter
3-4-1 model) that also undergoes a SSB up to $U(1)_{em}$. Its phenomenology
was exploited in a series of recent papers \cite{key-2} - \cite{key-4}
within the framework of the traditional approach. Apart from this,
we treat here a particular class of such models (namely, the one without
exotic electric charges) by resorting to the exact algebraical method
proposed more than a decade ago by Cot\u{a}escu \cite{key-5}. This
method, designed for gauge models with high symmetries with SSB, is
based on a proper minimal Higgs mechanism (mHm) employing a promissing
parametrization in the scalar sector, and consequently setting the
versors in the so called generalized Weinberg transformation (gWt)
that separates the electromagnetic field and diagonalizes the mass
matrix in the neutral (diagonal) bosons sector. As in the SM, only
one neutral scalar field finally remains. Its vacuum expectation value
(vev) $\left\langle \phi\right\rangle $ determines the overall breaking
scale of the model. 

We take for granted here the resulting formulas of the general method
and apply them to the particular 3-4-1 model of interest here. For
specific details of the method, the reader is referred to Ref \cite{key-5}.
We discriminate \cite{key-6} among he various classes of such models
by using the prescriptions suggested by the general method and we
avoid the classification carried out in Ref. \cite{key-4} which is
based on the well-known parameters $b$ and $c$. Notwithstanding,
we recover almost the same classes of models, except for the one treated
in Ref. \cite{key-7}. 

Our paper is organized as follows. Sec. 2 briefly reviews the main
features of the general method, while Sec. 3 presents the particle
content of the 3-4-1 models investigated in this paper and computes
both the mass terms in its boson sector and neutral charges of the
fermions involved therein. A one-parametr mass scale is finally given
for all the bosons. Sec. 3 presents our conclusions and phenomenological
estimates.

\section{The General Method}

\subsection{$SU(n)$ irreducible representations}

The general method mainly relies on the two fundamental irreducible
unitary representations (irreps) $\mathbf{n}$ and $\mathbf{n^{*}}$of
the $SU(n)$ group which are involved in contructing different classes
of tensors of ranks $(r,s)$ as direct products like $(\otimes{\textbf{n}})^{r}\otimes(\otimes{\textbf{n}}^{*})^{s}$.
These tensors have $r$ lower and $s$ upper indices for which we
reserve the notation, $i,j,k,\cdots=1,\cdots,n$. As usually, we denote
the irrep $\rho$ of $SU(n)$ by indicating its dimension, ${\mathbf{n}}_{\rho}$.
The $su(n)$ algebra can be parameterized in different ways, but here
it is convenient to use the hybrid basis of Ref. \cite{key-5} consisting
of $n-1$ diagonal generators of the Cartan subalgebra, $D_{\hat{i}}$,
labeled by indices $\hat{i},\hat{j},...$ ranging from $1$ to $n-1$,
and the generators $E_{j}^{i}=H_{j}^{i}/\sqrt{2}$, $i\not=j$, related
to the off-diagonal real generators $H_{j}^{i}$ \cite{key-8}. This
way the elements $\xi=D_{\hat{i}}\xi^{\hat{i}}+E_{j}^{i}\xi_{i}^{j}\in su(n)$
are now parameterized by $n-1$ real parameters, $\xi^{\hat{i}}$,
and by $n(n-1)/2$ $c$-number ones, $\xi_{j}^{i}=(\xi_{i}^{j})^{*}$,
for $i\not=j$. The advantage of this choice is that the parameters
$\xi_{j}^{i}$ can be directly associated to the $c$-number gauge
fields due to the factor $1/\sqrt{2}$ which gives their correct normalization.
In addition, this basis exhibit good trace orthogonality properties,
\begin{equation}
Tr(D_{\hat{i}}D_{\hat{j}})=\frac{1}{2}\delta_{\hat{i}\hat{j}},\quad Tr(D_{\hat{i}}E_{j}^{i})=0\,,\quad Tr(E_{j}^{i}E_{l}^{k})=\frac{1}{2}\delta_{l}^{i}\delta_{j}^{k}\,.\label{Eq.1}\end{equation}
 When we consider different irreps, $\rho$ of the $su(n)$ algebra
we denote $\xi^{\rho}=\rho(\xi)$ for each $\xi\in su(n)$ such that
the corresponding basis-generators of the irrep $\rho$ are $D_{\hat{i}}^{\rho}=\rho(D_{\hat{i}})$
and $E_{j}^{\rho\, i}=\rho(E_{j}^{i})$.

\subsection{Fermions}

The $U(1)_{Y}$ transformations are nothing else but phase factor
multiplications. Therefore - since the coupling constants $g$ for
$SU(n)_{L}$ and $g^{\prime}$ for the $U(1)_{Y}$ are assinged -
the transformation of the fermion tensor $L^{\rho}$ with respect
to the gauge group $SU(n)_{L}\otimes U(1)_{Y}$ of the theory reads
\begin{equation}
L^{\rho}\rightarrow U(\xi^{0},\xi)L^{\rho}=e^{-i(g\xi^{\rho}+g^{\prime}y_{ch}\xi^{0})}L^{\rho}\label{Eq.2}\end{equation}
 where $\xi=\in su(n)$ and $y_{ch}$ is the chiral hypercharge defining
the irrep of the $U(1)_{Y}$ group parametrized by $\xi^{0}$. For
simplicity, the general method deals with the character $y=y_{ch}g^{\prime}/g$
instead of the chiral hypercharge $y_{ch}$, but this mathematical
artifice does not affect in any way the results. Therefore, the irreps
of the whole gauge group $SU(n)_{L}\otimes U(1)_{Y}$ are uniquely
detemined by indicating the dimension of the $SU(n)$ tensor and its
character $y$ as $\rho=(\mathbf{n}_{\rho},y_{\rho})$.

In general, the spinor sector of our models has at least a part which
is put in pure left form using the charge conjugation. Consequently
this includes only left components, $L=\sum_{\rho}\oplus L^{\rho}$,
that transform according to an arbitrary reducible representation
of the gauge group. The Lagrangian density (Ld) of the free spinor
sector has the form \begin{equation}
{\mathcal{L}}_{S_{0}}=\frac{i}{2}\sum_{\rho}\overline{L^{\rho}}\tensor{\not\!\partial}L^{\rho}-\frac{1}{2}\sum_{\rho\rho'}\left(\overline{L^{\rho}}\chi^{\rho\rho^{\prime}}(L^{\rho^{\prime}})^{c}+h.c.\right).\label{Eq.3}\end{equation}
 Bearing in mind that each left-handed multiplet transforms as $L^{\rho}\rightarrow U^{\rho}(\xi^{0},\xi)L^{\rho}$
we understand that ${\mathcal{L}}_{S_{0}}$ remains invariant under
the global $SU(n)_{L}\otimes U(1)_{Y}$ transformations if the blocks
$\chi^{\rho\rho'}$ transform like $\chi^{\rho\rho^{\prime}}\rightarrow U^{\rho}(\xi^{0},\xi)\chi^{\rho\rho^{\prime}}(U^{\rho^{\prime}}(\xi^{0},\xi))^{T}$,
according to the representations $({\textbf{n}}_{\rho}\otimes{\textbf{n}}_{\rho'},y_{\rho}+y_{\rho'})$
which generally are reducible. These blocks will give rise to the
Yukawa terms.

\subsection{Gauge fields}

The spinor sector is coupled to the standard Yang-Mills sector constructed
in usual manner by gauging the $SU(n)_{L}\otimes U(1)_{Y}$ symmetry.
To this end we introduce the gauge fields $A_{\mu}^{0}=(A_{\mu}^{0})^{*}$
and $A_{\mu}=A_{\mu}^{+}=A_{\mu}^{a}T_{a}\in su(n)$. Furthermore,
the ordinary derivatives are replaced in Eq. (\ref{Eq.3}) by the
covariant ones, defined as $D_{\mu}L^{\rho}=\partial_{\mu}L^{\rho}-ig(A_{\mu}^{\rho}+y_{\rho}A_{\mu}^{0})L^{\rho}$.
Interaction terms occur

\subsection{Minimal Higgs mechanism}

The general method assumes also a particular Higgs mechanism (mHm)
based on a special parametrization in the scalar sector, such that
the $n$ Higgs multiplets $\phi^{(1)}$, $\phi^{(2)}$, ... $\phi^{(n)}$
satisfy the orthogonality condition $\phi^{(i)+}\phi^{(j)}=\phi^{2}\delta_{ij}$
in order to eliminate the unwanted Goldstone bosons that could survive
the SSB. $\phi$ is a gauge-invariant real scalar field while the
Higgs multiplets $\phi^{(i)}$ transform according to the irreps $(\mathbf{n},y^{(i)})$
whose characters $y^{(i)}$ are arbitrary numbers that can be organized
into the diagonal matrix $Y=Diag\left(y^{(1)},y^{(2)},\cdots,y^{(n)}\right)$.
The Higgs sector needs, in our approach, a parameter matrix \begin{equation}
\eta=Diag\left(\eta{}^{(1)},\eta{}^{(2)},...,\eta{}^{(n)}\right)\label{Eq.4}\end{equation}
 with the property ${\textrm{Tr}}(\eta^{2})=1-\eta_{0}^{2}$. It will
play the role of the metric in the kinetic part of the Higgs Ld which
reads \begin{equation}
\mathcal{L}_{H}=\frac{1}{2}\eta_{0}^{2}\partial_{\mu}\phi\partial^{\mu}\phi+\frac{1}{2}\sum_{i=1}^{n}\left(\eta{}^{(i)}\right)^{2}\left(D_{\mu}\phi^{(i)}\right)^{+}\left(D^{\mu}\phi^{(i)}\right)-V(\phi)\label{Eq.5}\end{equation}
 where $D_{\mu}\phi^{(i)}=\partial_{\mu}\phi^{(i)}-ig(A_{\mu}+y^{(i)}A_{\mu}^{0})\phi^{(i)}$
are the covariant derivatives of the model and $V(\phi)$ is the scalar
potential generating the SSB of the gauge symmetry \cite{key-5}.
This is assumed to have an absolute minimum for $\phi=\langle\phi\rangle\not=0$
that is, $\phi=\langle\phi\rangle+\sigma$ where $\sigma$ is the
unique surviving physical Higgs field. Therefore, one can always define
the unitary gauge where the Higgs multiplets, $\hat{\phi}^{(i)}$
have the components $\hat{\phi}_{k}^{(i)}=\delta_{ik}\phi=\delta_{ik}(\langle\phi\rangle+\sigma)$.

\subsection{Physical bosons}

The next step is to find the physical neutral bosons. Therefore, fist
one has to separate the electromagnetic potential $A_{\mu}^{em}$
corresponding to the surviving $U(1)_{em}$ symmetry. The one-dimensional
subspace of the parameters $\xi^{em}$ associated to this symmetry
assumes a particular direction in the parameter space $\lbrace\xi^{0},\xi^{\hat{i}}\rbrace$
of the whole Cartan subalgebra. This is uniquely determined by the
$n-1$ - dimensional unit vector $\nu$ and the angle $\theta$ giving
the subspace equations $\xi^{0}=\xi^{em}\cos\theta$ and $\xi^{\hat{i}}=\nu_{\hat{i}}\xi^{em}\sin\theta$.
On the other hand, since the Higgs multiplets in unitary gauge remain
invariant under $U(1)_{em}$ transformations, we must impose the obvious
condition $D_{\hat{i}}\xi^{\hat{i}}+Y\xi^{0}=0$ which yields $Y=-D_{\hat{i}}\nu^{\hat{i}}\tan\theta\equiv-(D\cdot\nu)\tan\theta$. 

In other words, the new parameters $(\nu,\theta)$ determine all the
characters $y^{(i)}$ of the irreps of the Higgs multiplets. For this
reason these will be considered the principal parameters of the model
and therefore one deals with $\theta$ and $\nu$ (which has $n-2$
independent components) instead of $n-1$ parameters $y^{(i)}$.

Under these circumstances, the generating mass term \begin{equation}
\frac{g^{2}}{2}\langle\phi\rangle^{2}Tr\left[\left(A_{\mu}+YA_{\mu}^{0}\right)\eta^{2}\left(A^{\mu}+YA^{0\mu}\right)\right]\,,\label{Eq.6}\end{equation}
 depends now on the parameters $\theta$ and $\nu_{\hat{i}}$. The
neutral bosons in Eq. \ref{Eq.6} being the electromagnetic field
$A_{\mu}^{em}$ and the $n-1$ new ones, $A_{\mu}^{'\hat{i}}$, which
are the diagonal bosons remaining after the separation of the electromagnetic
potential \cite{key-5}.

This term straightforwardly gives rise to the masses of the non-diagonal
gauge bosons \begin{equation}
M_{i}^{j}=\frac{1}{2}g\left\langle \phi\right\rangle \sqrt{\left[\left(\eta^{(i)}\right)^{2}+\left(\eta^{(j)}\right)^{2}\right]}\,,\label{Eq.7}\end{equation}
 while the masses of the neutral bosons $A_{\mu}^{'\hat{i}}$ have
to be calculated by diagonalizing the matrix \begin{equation}
(M^{2})_{\hat{i}\hat{j}}=\langle\phi\rangle^{2}Tr(B_{\hat{i}}B_{\hat{j}})\label{Eq.8}\end{equation}
 where \begin{equation}
B_{\hat{i}}=g\left(D_{\hat{i}}+\nu_{\hat{i}}(D\cdot\nu)\frac{1-\cos\theta}{\cos\theta}\right)\eta,\label{Eq.9}\end{equation}
 As it was expected, $A_{\mu}^{em}$ does not appear in the mass term
and, consequently, it remains massless. The other neutral gauge fields
${A'}_{\mu}^{\hat{i}}$ have the non-diagonal mass matrix (\ref{Eq.8}).
This can be brought in diagonal form with the help of a new $SO(n-1)$
transformation, $A_{\mu}^{'\hat{i}}=\omega_{\cdot\;\hat{j}}^{\hat{i}\;\cdot}Z_{\mu}^{\hat{j}}$
, which leads to the physical neutral bosons $Z_{\mu}^{\hat{i}}$
with well-defined masses. Performing this $SO(n-1)$ transformation
the physical neutral bosons are completely determined. The transformation
\begin{eqnarray}
A_{\mu}^{0} & = & A_{\mu}^{em}\cos\theta-\nu_{\hat{i}}\omega_{\cdot\;\hat{j}}^{\hat{i}\;\cdot}Z_{\mu}^{\hat{j}}\sin\theta,\label{Eq.10}\\
A_{\mu}^{\hat{k}} & = & \nu^{\hat{k}}A_{\mu}^{em}\sin\theta+\left(\delta_{\hat{i}}^{\hat{k}}-\nu^{\hat{k}}\nu_{\hat{i}}(1-\cos\theta)\right)\omega_{\cdot\;\hat{j}}^{\hat{i}\;\cdot}Z_{\mu}^{\hat{j}}.\label{Eq.11}\end{eqnarray}
 which switches from the original diagonal gauge fields, $(A_{\mu}^{0},A_{\mu}^{\hat{i}})$
to the physical ones, $(A_{\mu}^{em},Z_{\mu}^{\hat{i}})$ is called
the generalized Weinberg transformation (gWt).

\subsection{Electric and Neutral Charges}

The nest step is to identify the charges of the particles with the
coupling coefficients of the currents with respect to the above determined
physical bosons. Thus, we find that the spinor multiplet $L^{\rho}$
(of the irrep $\rho$) has the following electric charge matrix \begin{equation}
Q^{\rho}=g\left[(D^{\rho}\cdot\nu)\sin\theta+y_{\rho}\cos\theta\right],\label{Eq.12}\end{equation}
 and the $n-1$ neutral charge matrices \begin{equation}
Q^{\rho}(Z^{\hat{i}})=g\left[D_{\hat{k}}^{\rho}-\nu_{\hat{k}}(D^{\rho}\cdot\nu)(1-\cos\theta)-y_{\rho}\nu_{\hat{k}}\sin\theta\right]\omega_{\cdot\;\hat{i}}^{\hat{k}\;\cdot}\label{Eq.13}\end{equation}
 corresponding to the $n-1$ neutral physical fields, $Z_{\mu}^{\hat{i}}$.
All the other gauge fields, namely the charged bosons $A_{j\mu}^{i}$,
have the same coupling, $g/\sqrt{2}$, to the fermion multiplets.

\section{$SU(4)_{L}\otimes U(1)_{Y}$ models without exotic charges}

The general method - constructed in Ref. \cite{key-5} and briefly
presented in the above section - is based on the following assumptions
in order to give viable results when it is applied to concrete models:

({\small I}) the spinor sector must be put (at least partially) in
pure left form using the charge conjugation (see for details Appendix
B in Ref. \cite{key-5})

({\small II}) the minimal Higgs mechanism - with arbitrary parameters
$(\eta_{0},\eta)$ satisfying the condition ${\textrm{Tr}}(\eta^{2})=1-\eta_{0}^{2}$
and giving rise to traditional Yukawa couplings in unitary gauge -
must be employed

({\small III}) the coupling constant, $g$, is the same with the first
one of the SM

({\small IV}) at least one $Z$-like boson should satisfy the mass
condition $m_{Z}=m_{W}/\cos\theta_{W}$ established in the SM and
experimentally confirmed.

Bearing in mind all these necessary ingredients, we proceed to solving
the particular 3-4-1 model \cite{key-3} by imposing from the very
beginning the set of parameters we will work with. In the following,
we will use the standard generators $T_{a}$ of the $su(4)$ algebra.
Therefore, as the Hermitian diagonal generators of the Cartan subalgebra
one deals, in order, with $D_{1}=T_{3}=\frac{1}{2}Diag(1,-1,0,0)$,
$D_{2}=T_{8}=\frac{1}{2\sqrt{3}}Diag(1,1,-2,0)$, and $D_{3}=T_{15}=\frac{1}{2\sqrt{6}}Diag(1,1,1,-3)$
respectively. At the same time, we denote the irreps of the electroweak
model under consideration here by $\rho=(\mathbf{n}_{\rho},y_{ch}^{\rho})$
indicating the genuine chiral hypercharge $y_{ch}$ instead of $y$.
Therefore, the multiplets - subject to anomaly cancellation - of the
3-4-1 model of interest here will be denoted by $(\mathbf{n}_{color},\mathbf{n}_{\rho},y_{ch}^{\rho})$. 

There are three distinct cases \cite{key-6} leading to a discrimination
among models of the 3-4-1 class, according to their electric charge
assignment. They are: (i) versors $\nu_{1}=1$, $\nu_{2}=0$, $\nu_{3}=0$,
(ii) versors $\nu_{1}=0$, $\nu_{2}=1$, $\nu_{3}=0$, and (iii) versors
$\nu_{1}=0$, $\nu_{2}=0$, $\nu_{3}=-1$, respectively. At the same
time, one assumes the condition $e=g\sin\theta_{W}$ established in
the SM.

\subsection{Fermion content}

With this notation, after little algebra involving Eqs. (\ref{Eq.12})
- (\ref{Eq.14}) and the versor setting $\nu_{1}=0$, $\nu_{2}=0$,
$\nu_{3}=-1$ - Case 3 in ref. \cite{key-6} - one finds two distinct
classes of 3-4-1 models without exotic electric charges:First of all,
let's observe that no 4-plet obeys the fundamental irrep of the gauge
group $\rho=(\mathbf{4},0)$. Notwithstanding, since for the lepton
4-plet one can assign two different chiral hypercharges $-\frac{1}{4}$
and $-\frac{3}{4}$ respectively, we get two sub-cases leading to
two different versions of 3-4-1 anomaly-free models without exotic
electric charges. The coupling matching, as we will see in the following,
assumes the same relation in both sub-cases.

From Eq. (\ref{Eq.12}), it is straightforward that the lepton family
exhibits the electric charge operator

\begin{equation}
Q^{(4^{*},-\frac{1}{4})}=e\left[-T_{15}^{(4^{*})}\frac{\sin\theta}{\sin\theta_{W}}-\frac{1}{4}\left(\frac{g^{\prime}}{g}\right)\frac{\cos\theta}{\sin\theta_{W}}\right],\label{Eq.14}\end{equation}
for the first choice. This leads to the lepton representation $\left(N_{\alpha}^{\prime}\begin{array}{cccc}
, & N_{\alpha}, & \nu_{\alpha}, & e_{\alpha}\end{array}\right)_{L}^{T}\sim(\mathbf{4^{*}},-\frac{1}{4})$ including two new kinds of neutral leptons ($N_{\alpha}$, $N_{\alpha}^{\prime}$).
For the second choice, the electric charge operator will be represented
as 

\begin{equation}
Q^{(4,-\frac{1}{4})}=e\left[-T_{15}^{(4)}\frac{\sin\theta}{\sin\theta_{W}}-\frac{3}{4}\left(\frac{g^{\prime}}{g}\right)\frac{\cos\theta}{\sin\theta_{W}}\right],\label{Eq.15}\end{equation}
 leading to the lepton families $\left(E_{\alpha}^{\prime-}\begin{array}{cccc}
, & E_{\alpha}^{-}, & e_{\alpha}^{-}, & \nu_{\alpha}\end{array}\right)_{L}^{T}\sim(\mathbf{4},-\frac{3}{4})$ that allow for new charged leptons ($E_{\alpha}^{-}$, $E_{\alpha}^{\prime-}$).

After a little algebra, both Eqs (\ref{Eq.14}) and (\ref{Eq.15})
require - via the compulsory condition $\sin\theta=\sqrt{\frac{3}{2}}\sin\theta_{W}$,
since the only allowed electric charges in the lepton sector are $0$
and $\pm e$ - the coupling matching: $\frac{g^{\prime}}{g}=\frac{\sin\theta_{W}}{\sqrt{1-\frac{3}{2}\sin^{2}\theta_{W}}}$.

Once these assingments are assumed, the quarks will aquire their electric
charges from the following operators

\begin{equation}
Q^{(4^{*},\frac{5}{12})}=e\left[-T_{15}^{(4^{*})}\frac{\sin\theta}{\sin\theta_{W}}+\frac{5}{12}\left(\frac{g^{\prime}}{g}\right)\frac{\cos\theta}{\sin\theta_{W}}\right]\label{Eq.16}\end{equation}

\begin{equation}
Q^{(4,-\frac{1}{12})}=e\left[-T_{15}^{(4)}\frac{\sin\theta}{\sin\theta_{W}}-\frac{1}{12}\left(\frac{g^{\prime}}{g}\right)\frac{\cos\theta}{\sin\theta_{W}}\right]\label{Eq.17}\end{equation}

\subsubsection{Model A }

With the first of the above mentioned assumptions, the fermion representations
are:

\textbf{Lepton families}\begin{equation}
\begin{array}{ccccc}
f_{\alpha L}=\left(\begin{array}{c}
N_{\alpha}^{\prime}\\
N_{\alpha}\\
\nu_{\alpha}\\
e_{\alpha}\end{array}\right)_{L}\sim(\mathbf{1,4^{*}},-1/4) &  &  &  & \left(e_{\alpha L}\right)^{c}\sim(\mathbf{1,1},1)\end{array}\label{Eq.18}\end{equation}

\textbf{Quark families}\begin{equation}
\begin{array}{ccc}
Q_{iL}=\left(\begin{array}{c}
D_{i}^{\prime}\\
D_{i}\\
-d_{i}\\
u_{i}\end{array}\right)_{L}\sim(\mathbf{3,4},-1/12) &  & Q_{3L}=\left(\begin{array}{c}
U^{\prime}\\
U\\
u_{3}\\
d_{3}\end{array}\right)_{L}\sim(\mathbf{3},\mathbf{4^{*}},5/12)\end{array}\label{Eq.19}\end{equation}
\begin{equation}
\begin{array}{c}
(d_{3L})^{c},(d_{iL})^{c},(D_{iL})^{c},(D_{iL}^{\prime})^{c}\sim(\mathbf{3},\mathbf{1},+1/3)\end{array}\label{Eq.20}\end{equation}

\begin{equation}
(u_{3L})^{c},(u_{iL})^{c},(U_{L})^{c},(U_{L}^{\prime})^{c}\sim(\mathbf{3},\mathbf{1},-2/3)\label{Eq.21}\end{equation}
with $\alpha=1,2,3$ and $i=1,2$. We recovered the same fermion content
as the one of the model presented in Refs. \cite{key-3,key-4}.

\subsubsection{Model B}

With the second of the above mentioned assumptions, the fermion representations
are:

\textbf{Lepton families}\begin{equation}
\begin{array}{ccccc}
f_{\alpha L}=\left(\begin{array}{c}
E_{\alpha}^{\prime-}\\
E_{\alpha}^{-}\\
e_{\alpha}^{-}\\
e_{\alpha}\end{array}\right)_{L}\sim(\mathbf{1,4},-3/4) &  &  &  & \left(e_{\alpha L}\right)^{c},(E_{\alpha L})^{c},(E_{\alpha L}^{\prime})^{c}\sim(\mathbf{1,1},1)\end{array}\label{Eq.22}\end{equation}

\textbf{Quark families}\begin{equation}
\begin{array}{ccc}
Q_{iL}=\left(\begin{array}{c}
U_{i}^{\prime}\\
U_{i}\\
u_{i}\\
d_{i}\end{array}\right)_{L}\sim(\mathbf{3},\mathbf{4^{*}},5/12) &  & Q_{3L}=\left(\begin{array}{c}
D^{\prime}\\
D\\
-d_{3}\\
u_{3}\end{array}\right)_{L}\sim(\mathbf{3,4},-1/12)\end{array}\label{Eq.23}\end{equation}
\begin{equation}
\begin{array}{c}
(d_{3L})^{c},(d_{iL})^{c},(D_{iL})^{c},(D_{iL}^{\prime})^{c}\sim(\mathbf{3},\mathbf{1},+1/3)\end{array}\label{Eq.24}\end{equation}

\begin{equation}
(u_{3L})^{c},(u_{iL})^{c},(U_{L})^{c},(U_{L}^{\prime})^{c}\sim(\mathbf{3},\mathbf{1},-2/3)\label{Eq.25}\end{equation}
with $\alpha=1,2,3$ and $i=1,2$. We recovered the same fermion content
as the one of the model presented in Refs. \cite{key-3,key-4}.

With this assignment the fermion families (in each of the above displayed
cases) cancel the axial anomalies by just an interplay between them,
although each family remains anomalous by itself. Thus, the renormalization
criteria are fulfilled and the method is validated once more from
this point of view. Note that one can add at any time sterile neutrinos
- \emph{i.e.} right-handed neutrinos $\nu_{\alpha R}\sim(\mathbf{1,1},0)$
- that could pair in the neutrino sector of the Ld with left-handed
ones in order to eventually generate tiny Dirac or Majorana masses
by means of an adequate see-saw mechanism. These sterile neutrinos
do not affect anyhow the anomaly cancelation, since all their charges
are zero. Moreover, their number is not restricted by the number of
flavors in the model

\subsection{Boson mass spectrum}

Subsequently, we will use the standard generators $T_{a}$ of the
$su(4)$ algebra. In this basis, the gauge fields are $A_{\mu}^{0}$
and $A_{\mu}\in su(4)$, that is \begin{equation}
A_{\mu}=\frac{1}{2}\left(\begin{array}{ccccccc}
D_{\mu}^{1} &  & \sqrt{2}Y_{\mu} &  & \sqrt{2}X_{\mu}^{\prime} &  & \sqrt{2}X_{\mu}^{\prime}\\
\\\sqrt{2}Y_{\mu}^{*} &  & D_{\mu}^{2} &  & \sqrt{2}K_{\mu} &  & \sqrt{2}K_{\mu}^{\prime}\\
\\\sqrt{2}X_{\mu}^{*}{} &  & \sqrt{2}K_{\mu}^{*} &  & D_{\mu}^{3} &  & \sqrt{2}W_{\mu}\\
\\\sqrt{2}X_{\mu}^{\prime*} &  & \sqrt{2}K_{\mu}^{\prime*} &  & \sqrt{2}W_{\mu}^{*} &  & D_{\mu}^{4}\end{array}\right),\label{Eq.26}\end{equation}
with $D_{\mu}^{1}=A_{\mu}^{3}+A_{\mu}^{8}/\sqrt{3}+A_{\mu}^{15}/\sqrt{6}$,
$D_{\mu}^{2}=-A_{\mu}^{3}+A_{\mu}^{8}/\sqrt{3}+A_{\mu}^{15}/\sqrt{6}$,
$D_{\mu}^{3}=-2A_{\mu}^{8}/\sqrt{3}+A_{\mu}^{15}/\sqrt{6}$, $D_{\mu}^{4}=-3A_{\mu}^{15}/\sqrt{6}$
as diagonal bosons. Apart from the charged Weinberg bosons ($W^{\pm}$),
there are two new charged bosons, $K^{0}$, $K^{\prime\pm}$, while
$X^{0}$, $X^{\prime\pm}$ and $Y^{0}$ are new neutral bosons, but
distinct from the diagonl ones.

The masses of both the neutral and charged bosons depend on the choice
of the matrix $\eta$ whose components are free parameters. Here it
is convenient to assume the following matrix \begin{equation}
\eta^{2}=(1-\eta_{0}^{2})Diag\left(1-c,c-a,\frac{1}{2}a+b,\frac{1}{2}a-b\right),\label{Eq.27}\end{equation}
 where, for the moment, $a$,$b$ and $c$ are arbitrary non-vanishing
real parameters. Obviously, $\eta_{0},c\in[0,1)$, $a\in(0,c)$ and
$b\in(-a,+a)$. Note that with this parameter choice the condition
(II) is accomplished. 

Under these circumstances, the mass spectrum corresponding to the
off-diagonal bosons (usually, charged ones), according to Eq. (\ref{Eq.7})
reads \begin{eqnarray}
{m}^{2}(W) & = & m^{2}a,\label{Eq.28}\\
{m}^{2}(X) & = & m^{2}\left(1-c+\frac{{\textstyle 1}}{2}a+b\right),\label{Eq.29}\\
{m}^{2}(X') & = & m^{2}\left(1-c+\frac{{\textstyle 1}}{2}a-b\right),\label{Eq.30}\\
{m}^{2}(K) & ={} & m^{2}\left(c-\frac{{\textstyle 1}}{2}a+b\right),\label{Eq.31}\\
{m}^{2}(K') & = & m^{2}\left(c-\frac{{\textstyle 1}}{2}a-b\right),\label{Eq.32}\\
{m}^{2}(Y) & = & m^{2}\left(1-a\right).\label{Eq.33}\end{eqnarray}
 while the mass matrix of the neutral bosons is given by Eq. (\ref{Eq.8})
\begin{equation}
M^{2}=m^{2}\left(\begin{array}{ccccc}
\left(1-a\right) &  & \frac{{\textstyle 1-2c+a}}{{\textstyle \sqrt{3}}} &  & \frac{{\textstyle 1-2c+a}}{{\textstyle \sqrt{6}\cos\theta}}\\
\\\frac{{\textstyle 1-2c+a}}{{\textstyle \sqrt{3}}} &  & \frac{{\textstyle 1}}{{\textstyle 3}}\left(1+a+4b\right) &  & \frac{{\textstyle 1-2a-2b}}{{\textstyle 3\sqrt{2}\cos\theta}}\\
\\\frac{{\textstyle 1-2c+a}}{{\textstyle \sqrt{6}\cos\theta}} &  & \frac{{\textstyle 1-2a-2b}}{{\textstyle 3\sqrt{2}\cos\theta}} &  & \frac{{\textstyle 1+4a-8b}}{6{\textstyle \cos^{2}\theta}}\end{array}\right)\label{Eq.34}\end{equation}
 with $m^{2}=g^{2}\left\langle \phi\right\rangle ^{2}(1-\eta_{0}^{2})/4$
throughout this paper. In order to fulfil the requirement (IV), the
above matrix has to admit $m^{2}a/{\textstyle \cos^{2}\theta}_{W}$
as eigenvalue, that is one has to compute $Det\left|M^{2}-\frac{m^{2}a}{\cos^{2}\theta_{W}}\right|=0$.
Now, one can enforce some other phenomenological assumptions. First
of all, it is natural to presume that the third neutral (diagonal
boson) $Z^{\prime\prime}$should be considered much heavier than its
companions, so that it decouples form their mixing, as the symmetry
is broken to $SU(3)$. For this purpose a higher breaking scale is
responsable. 

Therefore $M_{12}^{2}=M_{21}^{2}=M_{13}^{2}=M_{31}^{2}$ in the matrix
(\ref{Eq.34}). This gives rise to the natural condition

\begin{equation}
c=\frac{1+a}{2},\label{Eq.35}\end{equation}
in order to vanish the above terms. 

Hence, Eq. (\ref{Eq.34}) looks like

\begin{equation}
M^{2}=m^{2}\left(\begin{array}{ccc}
(1-a) & 0 & 0\\
\\0 & \frac{1+a+4b}{{\textstyle 3}} & \frac{1-2a-2b}{{\textstyle 3\sqrt{2}\cos\theta}}\\
\\0 & \frac{1-2a-2b}{{\textstyle 3\sqrt{2}\cos\theta}} & \frac{1+4a-8b}{6{\textstyle \cos^{2}\theta}}\end{array}\right)\label{Eq.36}\end{equation}

Let us observe that the condition (IV) - via computing $Det\left|M^{2}-\frac{m^{2}a}{\cos^{2}\theta_{W}}\right|=0$
- is fulfilled if and only if $b=\frac{1}{2}a\tan^{2}\theta_{W}$,
resulting from diagonalization of the remaning part of the matrix
(\ref{Eq.36}). Therefore, one finally remains with only one parameter
- say $a$. 

Obviously, $Z$ is the neutral boson of the SM, while $Z^{\prime}$
is a new neutral boson of this model (also occuring in 3-3-1 models)
whose mass comes form $Tr(M^{2})={m}^{2}(Z)+{m}^{2}(Z^{\prime})+{m}^{2}(Z^{\prime\prime})$. 

With these preliminaries, the boson mass spectrum holds:

\begin{eqnarray}
{m}^{2}(W) & = & m^{2}a,\label{Eq.37}\\
{m}^{2}(X) & = & m^{2}a\left(\frac{1+\tan^{2}\theta_{W}}{2}\right),\label{Eq.38}\\
{m}^{2}(X') & = & m^{2}a\left(\frac{1-\tan^{2}\theta_{W}}{2}\right),\label{Eq.39}\\
{m}^{2}(K) & ={} & m^{2}a\left(\frac{1+\tan^{2}\theta_{W}}{2}\right),\label{Eq.40}\\
{m}^{2}(K') & = & m^{2}a\left(\frac{1-\tan^{2}\theta_{W}}{2}\right),\label{Eq.41}\\
{m}^{2}(Y) & = & m^{2}(1-a),\label{Eq.42}\\
{m}^{2}(Z) & = & m^{2}a/\cos^{2}\theta_{W},\label{Eq.43}\\
{m}^{2}(Z^{\prime}) & = & {m}^{2}\frac{\cos^{4}\theta_{W}-a\sin^{4}\theta_{W}}{\cos^{2}\theta_{W}\left(2-3\sin^{2}\theta_{W}\right)},\label{Eq.44}\\
{m}^{2}(Z^{\prime\prime}) & = & m^{2}(1-a).\label{Eq.45}\end{eqnarray}

The mass scale is now just a matter of tuning the parameter $a$ in
accordance with the possible values for $\left\langle \phi\right\rangle $.

\subsection{Neutral charges}

Now one can compute in detail all the charges for the fermion representations
in models A and B with respect to the neutral bosons ($Z$, $Z^{\prime}$,
$Z^{\prime\prime}$), since the gWt is determined by the matrix

\begin{equation}
\omega=\left(\begin{array}{ccccc}
1 &  & 0 &  & 0\\
\\0 &  & \frac{1}{\sqrt{3}\sqrt{1-\sin^{2}\theta_{W}}} &  & \frac{\sqrt{2-3\sin^{2}\theta_{W}}}{\sqrt{3}\sqrt{1-\sin^{2}\theta_{W}}}\\
\\0 &  & -\frac{\sqrt{2-3\sin^{2}\theta_{W}}}{\sqrt{3}\sqrt{1-\sin^{2}\theta_{W}}} &  & \frac{1}{\sqrt{3}\sqrt{1-\sin^{2}\theta_{W}}}\end{array}\right).\label{Eq.46}\end{equation}

These will be expressed - via Eq.(\ref{Eq.13}) with the versor assignment
$\nu_{1}=0$, $\nu_{2}=0$, $\nu_{3}=-1$ - by:

\begin{equation}
Q^{\rho}(Z^{\hat{i}})=g\left[D_{1}^{\rho}\omega_{\cdot\;\hat{i}}^{1\;\cdot}+D_{2}^{\rho}\omega_{\cdot\;\hat{i}}^{2\;\cdot}+\left(D_{3}^{\rho}\cos\theta+y_{ch}^{\rho}\frac{g^{\prime}}{g}\sin\theta\right)\omega_{\cdot\;\hat{i}}^{3\;\cdot}\right],\label{Eq.47}\end{equation}
where the conditions $\frac{g^{\prime}}{g}=\frac{\sin\theta_{W}}{\sqrt{1-\frac{3}{2}\sin^{2}\theta_{W}}}$
and $\sin\theta=\sqrt{\frac{3}{2}}\sin\theta_{W}$ have to be inserted.

\begin{table}

\caption{Coupling coefficients of the neutral currents in 3-4-1 model A}

\begin{tabular}{cccccc}
\hline 
Particle\textbackslash{}Coupling($e/\sin2\theta_{W}$)&
$Z\rightarrow\bar{f}f$&
&
$Z^{\prime}\rightarrow\bar{f}f$&
&
$Z^{\prime\prime}\rightarrow\bar{f}f$\tabularnewline
\hline
\hline 
&
&
&
&
&
\tabularnewline
&
&
&
&
&
\tabularnewline
$\nu_{eL},\nu_{\mu L},\nu_{\tau L}$&
$1$&
&
$\frac{1-3\sin^{2}\theta_{W}}{2\sqrt{2-3\sin^{2}\theta_{W}}}$&
&
$0$\tabularnewline
&
&
&
&
&
\tabularnewline
$e_{L},\mu_{L},\tau_{L}$&
$2\sin^{2}\theta_{W}-1$&
&
$\frac{1-3\sin^{2}\theta_{W}}{2\sqrt{2-3\sin^{2}\theta_{W}}}$&
&
$0$\tabularnewline
&
&
&
&
&
\tabularnewline
$N_{eL},N_{\mu L},N_{\tau L}$&
$0$&
&
$-\frac{3\cos^{2}\theta_{W}}{2\sqrt{2-3\sin^{2}\theta_{W}}}$&
&
$\cos\theta_{W}$\tabularnewline
&
&
&
&
&
\tabularnewline
$N_{eL}^{\prime},N_{\mu L}^{\prime},N_{\tau L}^{\prime}$&
$0$&
&
$-\frac{3\cos^{2}\theta_{W}}{2\sqrt{2-3\sin^{2}\theta_{W}}}$&
&
$-\cos\theta_{W}$\tabularnewline
&
&
&
&
&
\tabularnewline
$e_{R},\mu_{R},\tau_{R}$&
$2\sin^{2}\theta_{W}$&
&
$-\frac{2\sin^{2}\theta_{W}}{\sqrt{2-3\sin^{2}\theta_{W}}}$&
&
$0$\tabularnewline
&
&
&
&
&
\tabularnewline
$u_{L},c_{L}$&
$1-\frac{4}{3}\sin^{2}\theta_{W}$&
&
$\frac{2-9\cos^{2}\theta_{W}}{2\sqrt{2-3\sin^{2}\theta_{W}}}$&
&
$0$\tabularnewline
&
&
&
&
&
\tabularnewline
$d_{L},s_{L}$&
$-1+\frac{2}{3}\sin^{2}\theta_{W}$&
&
$\frac{2-9\cos^{2}\theta_{W}}{2\sqrt{2-3\sin^{2}\theta_{W}}}$&
&
0\tabularnewline
&
&
&
&
&
\tabularnewline
$t_{L}$&
$1-\frac{4}{3}\sin^{2}\theta_{W}$&
&
$\frac{2+9\cos^{2}\theta_{W}}{6\sqrt{2-3\sin^{2}\theta_{W}}}$&
&
0\tabularnewline
&
&
&
&
&
\tabularnewline
$b_{L}$&
$-1+\frac{2}{3}\sin^{2}\theta_{W}$&
&
$\frac{2+9\cos^{2}\theta_{W}}{6\sqrt{2-3\sin^{2}\theta_{W}}}$&
&
0\tabularnewline
&
&
&
&
&
\tabularnewline
$u_{R},c_{R},t_{R},U_{1R},U_{iR}^{\prime}$&
$-\frac{4}{3}\sin^{2}\theta_{W}$&
&
$\frac{4\sin^{2}\theta_{W}}{3\sqrt{2-3\sin^{2}\theta_{W}}}$&
&
$0$\tabularnewline
&
&
&
&
&
\tabularnewline
$d_{R},s_{R},b_{R},D_{iR},D_{iR}^{\prime}$&
$+\frac{2}{3}\sin^{2}\theta_{W}$&
&
$-\frac{2\sin^{2}\theta_{W}}{3\sqrt{2-3\sin^{2}\theta_{W}}}$&
&
$0$\tabularnewline
&
&
&
&
&
\tabularnewline
$D_{1L},D_{2L}$&
$\frac{2}{3}\sin^{2}\theta_{W}$&
&
$\frac{5-9\sin^{2}\theta_{W}}{6\sqrt{2-3\sin^{2}\theta_{W}}}$&
&
$-\cos\theta_{W}$\tabularnewline
&
&
&
&
&
\tabularnewline
$D_{1L}^{\prime},D_{2L}^{\prime}$&
$\frac{2}{3}\sin^{2}\theta_{W}$&
&
$\frac{5-9\sin^{2}\theta_{W}}{6\sqrt{2-3\sin^{2}\theta_{W}}}$&
&
$\cos\theta_{W}$\tabularnewline
&
&
&
&
&
\tabularnewline
$U_{3L}$&
$-\frac{4}{3}\sin^{2}\theta_{W}$&
&
$\frac{-1+9\sin^{2}\theta_{W}}{6\sqrt{2-3\sin^{2}\theta_{W}}}$&
&
$\cos\theta_{W}$\tabularnewline
&
&
&
&
&
\tabularnewline
$U_{3L}^{\prime}$&
$-\frac{4}{3}\sin^{2}\theta_{W}$&
&
$\frac{-1+9\sin^{2}\theta_{W}}{6\sqrt{2-3\sin^{2}\theta_{W}}}$&
&
$-\cos\theta_{W}$\tabularnewline
&
&
&
&
&
\tabularnewline
&
&
&
&
&
\tabularnewline
\hline 
&
&
&
&
&
\tabularnewline
\end{tabular}
\end{table}

\begin{table}

\caption{Coupling coefficients of the neutral currents in 3-4-1 Model B}

\begin{tabular}{cccccc}
\hline 
Particle\textbackslash{}Coupling($e/\sin2\theta_{W}$)&
$Z\rightarrow\bar{f}f$&
&
$Z^{\prime}\rightarrow\bar{f}f$&
&
$Z^{\prime\prime}\rightarrow\bar{f}f$\tabularnewline
\hline
\hline 
&
&
&
&
&
\tabularnewline
&
&
&
&
&
\tabularnewline
$\nu_{eL},\nu_{\mu L},\nu_{\tau L}$&
$1$&
&
$\frac{-5+3\sin^{2}\theta_{W}}{2\sqrt{2-3\sin^{2}\theta_{W}}}$&
&
$0$\tabularnewline
&
&
&
&
&
\tabularnewline
$e_{L},\mu_{L},\tau_{L}$&
$2\sin^{2}\theta_{W}-1$&
&
$\frac{-5+3\sin^{2}\theta_{W}}{2\sqrt{2-3\sin^{2}\theta_{W}}}$&
&
$0$\tabularnewline
&
&
&
&
&
\tabularnewline
$E_{eL},E_{\mu L},E_{\tau L}$&
$2\sin^{2}\theta_{W}$&
&
$-\frac{1+3\sin^{2}\theta_{W}}{2\sqrt{2-3\sin^{2}\theta_{W}}}$&
&
$-\cos\theta_{W}$\tabularnewline
&
&
&
&
&
\tabularnewline
$E_{eL}^{\prime},E_{\mu L}^{\prime},E_{\tau L}^{\prime}$&
$2\sin^{2}\theta_{W}$&
&
$-\frac{1+3\sin^{2}\theta_{W}}{2\sqrt{2-3\sin^{2}\theta_{W}}}$&
&
$\cos\theta_{W}$\tabularnewline
&
&
&
&
&
\tabularnewline
$e_{iR},E_{iR},E_{iR}^{\prime}$&
$2\sin^{2}\theta_{W}$&
&
$-\frac{2\sin^{2}\theta_{W}}{\sqrt{2-3\sin^{2}\theta_{W}}}$&
&
$0$\tabularnewline
&
&
&
&
&
\tabularnewline
$u_{L},c_{L}$&
$1-\frac{4}{3}\sin^{2}\theta_{W}$&
&
$\frac{2+9\cos^{2}\theta_{W}}{6\sqrt{2-3\sin^{2}\theta_{W}}}$&
&
$0$\tabularnewline
&
&
&
&
&
\tabularnewline
$d_{L},s_{L}$&
$-1+\frac{2}{3}\sin^{2}\theta_{W}$&
&
$\frac{2+9\cos^{2}\theta_{W}}{6\sqrt{2-3\sin^{2}\theta_{W}}}$&
&
0\tabularnewline
&
&
&
&
&
\tabularnewline
$t_{L}$&
$1-\frac{4}{3}\sin^{2}\theta_{W}$&
&
$\frac{2-9\cos^{2}\theta_{W}}{2\sqrt{2-3\sin^{2}\theta_{W}}}$&
&
0\tabularnewline
&
&
&
&
&
\tabularnewline
$b_{L}$&
$-1+\frac{2}{3}\sin^{2}\theta_{W}$&
&
$\frac{2-9\cos^{2}\theta_{W}}{2\sqrt{2-3\sin^{2}\theta_{W}}}$&
&
0\tabularnewline
&
&
&
&
&
\tabularnewline
$u_{R},c_{R},t_{R},U_{1R},U_{iR}^{\prime}$&
$-\frac{4}{3}\sin^{2}\theta_{W}$&
&
$\frac{4\sin^{2}\theta_{W}}{3\sqrt{2-3\sin^{2}\theta_{W}}}$&
&
$0$\tabularnewline
&
&
&
&
&
\tabularnewline
$d_{R},s_{R},b_{R},D_{iR},D_{iR}^{\prime}$&
$+\frac{2}{3}\sin^{2}\theta_{W}$&
&
$-\frac{2\sin^{2}\theta_{W}}{3\sqrt{2-3\sin^{2}\theta_{W}}}$&
&
$0$\tabularnewline
&
&
&
&
&
\tabularnewline
$D_{3L}$&
$\frac{2}{3}\sin^{2}\theta_{W}$&
&
$\frac{5-9\sin^{2}\theta_{W}}{6\sqrt{2-3\sin^{2}\theta_{W}}}$&
&
$-\cos\theta_{W}$\tabularnewline
&
&
&
&
&
\tabularnewline
$D_{3L}^{\prime}{}$&
$\frac{2}{3}\sin^{2}\theta_{W}$&
&
$\frac{5-9\sin^{2}\theta_{W}}{6\sqrt{2-3\sin^{2}\theta_{W}}}$&
&
$\cos\theta_{W}$\tabularnewline
&
&
&
&
&
\tabularnewline
$U_{1L},U_{2L}$&
$-\frac{4}{3}\sin^{2}\theta_{W}$&
&
$\frac{-1+9\sin^{2}\theta_{W}}{6\sqrt{2-3\sin^{2}\theta_{W}}}$&
&
$\cos\theta_{W}$\tabularnewline
&
&
&
&
&
\tabularnewline
$U_{1L}^{\prime},U_{2L}^{\prime}$&
$-\frac{4}{3}\sin^{2}\theta_{W}$&
&
$\frac{-1+9\sin^{2}\theta_{W}}{6\sqrt{2-3\sin^{2}\theta_{W}}}$&
&
$-\cos\theta_{W}$\tabularnewline
&
&
&
&
&
\tabularnewline
&
&
&
&
&
\tabularnewline
\hline 
&
&
&
&
&
\tabularnewline
\end{tabular}
\end{table}

Evidently, the heaviest neutral boson - $Z^{1}=Z^{\prime\prime}$,
in our notation - will couple the fermion representations through:

\begin{equation}
Q^{\rho}(Z^{1})=gD_{1}^{\rho}\label{Eq.48}\end{equation}
while the other two - $Z^{2}=Z$of the SM, and $Z^{3}=Z^{\prime}$respectively
- exhibit the following charges:

\begin{equation}
Q^{\rho}(Z^{2})=g\left[D_{2}^{\rho}\omega_{\cdot\;2}^{2\;\cdot}+\left(D_{3}^{\rho}\sqrt{1-\frac{3}{2}\sin^{2}\theta_{W}}+y_{ch}^{\rho}\frac{\sqrt{3}\sin^{2}\theta_{W}}{\sqrt{2-3\sin^{2}\theta_{W}}}\right)\omega_{\cdot\;2}^{3\;\cdot}\right]\label{Eq.49}\end{equation}

\begin{equation}
Q^{\rho}(Z^{3})=g\left[D_{2}^{\rho}\omega_{\cdot\;3}^{2\;\cdot}+\left(D_{3}^{\rho}\sqrt{1-\frac{3}{2}\sin^{2}\theta_{W}}+y_{ch}^{\rho}\frac{\sqrt{3}\sin^{2}\theta_{W}}{\sqrt{2-3\sin^{2}\theta_{W}}}\right)\omega_{\cdot\;3}^{3\;\cdot}\right]\label{Eq.50}\end{equation}

Assuming the $\omega$- matrix given by Eq. (\ref{Eq.46}), the neutral
charges of the fermions in the two models under consideration here
are computed and listed in Tables 1 and 2.

\section{Conclusions}

Regarding the neutral currents, one can observe that the leptons and
quarks of the SM recover their known values with respect to the $Z$
boson, while each exotic fermion in the 3-4-1 models of interest here
exhibit a vector coupling with respect to the same boson, \emph{i.e.}
its left-handed component and its right-handed one are indistinct
in interaction with $Z$. On the other hand, $Z^{\prime\prime}$ couples
only the exotic fermions. 

In order to allow for a high breaking scale in the model $\langle\phi\rangle\geq1$TeV
and keep at the same time consistency with low energy phenomenology
of the SM our solution favors the case with $a\rightarrow0$ and $c\rightarrow\frac{1}{2}$.
However, assuming that $m(W)\simeq84.4$GeV and $m(Z)\simeq91.2$GeV
and $\sin^{2}\theta_{W}\simeq0.223$ \cite{key-9}, our approach predicts
the exact masses at tree level for the following bosons (according
to Eqs. (\ref{Eq.37}) - (\ref{Eq.45})): $m(X)\equiv m(K)\simeq67.7$GeV,
$m(X^{\prime})\equiv m(K^{\prime})\simeq50.4$GeV which are independent
of the precise account of the overall vev of the model. The heavier
bosons - essentially depending on the precise account of the vev,
and hence on the parameter $a$- fulfil the following hierarchy $m(Z^{\prime\prime})\equiv m(Y)>m(Z^{\prime})>m(Z)>m(W)>m(X)\equiv m(K)>m(X^{\prime})\equiv m(K^{\prime})$. 

We can offer here a rough estimate. If the mass scale of the model
($m$ in our notation), lies in the TeV region or in a higher one,
then from Eqs. \ref{Eq.37} , with $m\simeq1$TeV, $a\simeq0.007$
is inferred. Hence, by working out Eqs. (\ref{Eq.42}) (\ref{Eq.44})
and (\ref{Eq.45}) the rest of the bosons are given their masses accordingly:
$m(Z^{\prime\prime})\equiv m(Y)\simeq0.996$TeV and $m(Z^{\prime})\simeq0.55$TeV.
A more accurate estimate for the masses of these bosons and the relations
among them (by a more apropriate tuning of parameter $a$) can be
done, once the experimental evidence of their phenomenology will be
definitely bring to light at LHC, LEP, CDF, Tevatron and other high
energy accelerators in a near future.

\end{document}